\documentclass[aps,prl,twocolumn,groupedaddress,nofootinbib]{revtex4}
\usepackage[pass,letterpaper]{geometry}

\pdfoutput=1  
\usepackage{bm}
\usepackage{graphicx,color}
\usepackage{latexsym,amsmath,amssymb,graphicx,booktabs}
\usepackage{hyperref}
\usepackage{placeins}
\usepackage{subfigure}
\usepackage{mathtools}
\usepackage[normalem]{ulem}

\definecolor{MyBlue}{rgb}{0.15,0.15,0.70}
\definecolor{Dgreen}{rgb}{0,0.7,0.0}

\hypersetup{
colorlinks=true,
citecolor=MyBlue,
linkcolor=MyBlue,
urlcolor=MyBlue
}

\usepackage{amssymb}
\usepackage{amsmath}
\usepackage{amsfonts}
\usepackage{upgreek}
\usepackage{latexsym}
\usepackage{appendix}
\usepackage{eufrak}
\usepackage{dsfont}

\newcommand\ees{\end{eqnarray}}
\newcommand\bees{\begin{eqnarray}}

\usepackage[export]{adjustbox}

\newcommand\spart{\;\raise1.0pt\hbox{/}\hskip-6pt\partial}
\newcommand\spartb{\;\overline{\raise1.0pt\hbox{/}\hskip-6pt\partial}}

\newcommand{\be}{\begin{equation}}
\newcommand{\ee}{\end{equation}}
\newcommand{\bk}{{\bm k}}

\newcommand{\beqa}{\begin{eqnarray}}
\newcommand{\eeqa}{\end{eqnarray}}

\newcommand{\bee}{{\bm{e}}}

\newcommand{\m}{\rm{m}}

\newcommand{\dd}{\text{d}}

\newcommand{\nn}{\nonumber}

\newcommand{\obs}{_{\rm O}}
\newcommand{\Gal}{_{\rm G}}

\newcommand{\GW}{{_{\rm GW}}}

\begin{document}
\title{First predictions of the angular power spectrum of the astrophysical
    gravitational wave background}
\author{Giulia Cusin}
\email{giulia.cusin@physics.ox.ac.uk}
\affiliation{Astrophysics Department, University of Oxford, DWB, Keble Road, Oxford OX1 3RH, UK}
\author{Irina Dvorkin}
\affiliation{Max Planck Institute for Gravitational Physics (Albert Einstein Institute), Am M\"{u}hlenberg 1, Potsdam-Golm, 14476, Germany}
\author{Cyril Pitrou}
\affiliation{Institut d'Astrophysique de Paris, CNRS UMR 7095, 
\\
           Sorbonne Universit\'e, Institut Lagrange de Paris, 98 bis, Bd Arago, 75014 Paris, France}
\author{Jean-Philippe Uzan}
\affiliation{Institut d'Astrophysique de Paris, CNRS UMR 7095, 
\\
           Sorbonne Universit\'e, Institut Lagrange de Paris, 98 bis, Bd Arago, 75014 Paris, France}
\vspace{1 em}
\date{\today}

\begin{abstract}
We present the first predictions for the angular power spectrum of the astrophysical gravitational wave background constituted of the radiation emitted by all resolved and unresolved astrophysical sources. Its shape and amplitude depend on both the astrophysical properties on galactic scales and on cosmological properties.  We show that the angular power spectrum behaves as $C_\ell \propto 1/\ell$ on large scales and that relative fluctuations of the signal are of order 30\% at $100\,{\rm Hz}$. We also present the correlations of the astrophysical gravitational wave background with weak-lensing and galaxy distribution. These numerical results pave the way to the study of a new observable at the crossroad between general relativity, astrophysics and cosmology.
\end{abstract}
\maketitle

\noindent{\bf Introduction.} The detection of the first gravitational wave (GW) source~\cite{Abbott:2016blz} by the Advanced Laser Interferometric Gravitational-wave Observatory (LIGO) has triggered the birth of a new window in astronomy. \textcolor{black}{Gravitational-wave} astronomy has so far mostly focused on the study of resolved sources. However, complementary information may come also from the study of the superposition of signals from unresolved astrophysical sources.

Indeed, astronomical observations  include diffuse stochastic backgrounds of radiations due to the superposition of the signals from all, resolved and unresolved, sources. The electromagnetic backgrounds include the cosmic microwave background (CMB) with a black body spectrum~\cite{1965ApJ...142..419P} and the extragalactic background made up of all the electromagnetic radiation emitted by stars, galaxies, galaxy clusters etc. since their formation~\cite[e.g.][]{1967ApJ...148..377P,1991Natur.353..315S,2001ARA&A..39..249H}. Similarly, there exists a neutrino background~\cite{2006ARNPS..56..137H} and indeed a GW background.  

This latter has several components contributing at different frequencies and with different statistical properties~\cite{Regimbau:2011rp, Zhu:2011bd}; its spectrum is defined by the dimensionless density parameter
\begin{equation}\label{e.Ogwdef}
  \bar\Omega_{\GW}(\nu_{\obs})=\frac{\nu_{\obs}}{\rho_c}\frac{\dd\rho_\GW}{\dd\nu_{\obs}}\,,
\end{equation}
where $\rho_c=3H_0^2/8\pi G$ is the critical density of the Universe and $\nu_{\obs}$ is the frequency measured by the observer. In standard cosmology, one expects a primordial component of GW produced during inflation but also during preheating~\cite{Dufaux:2007pt}. Many other more speculative sources have been considered; see e.g.  Refs.~\cite{Binetruy:2012ze, Buonanno:2014aza}. In addition, the astrophysical GW background (AGWB) contribution  arises from the superposition of a large number of unresolved astrophysical sources since the beginning of stellar activity. 
 
This letter focuses on this last component of the AGWB to which many sources may contribute such as merging binary black holes (BH) and neutron stars (NS)~\cite{Regimbau:2016ike, Mandic:2016lcn, Dvorkin:2016okx, Nakazato:2016nkj, Dvorkin:2016wac, Evangelista:2014oba}, supermassive binary BH mergers~\cite{2003ApJ...590..691W,2008MNRAS.390..192S,2016PhRvD..93b4003K,2017MNRAS.470.4547D,Kelley:2017lek}, rotating NS~\cite{Surace:2015ppq, Talukder:2014eba, Lasky:2013jfa}, stellar core collapses~\cite{Crocker:2015taa,Crocker:2017agi}. Its properties depend on (1) local astrophysics through the energy spectrum of each kind of sources, (2) the host galaxies through the abundance of these systems and their evolution and (3) cosmology through the distribution of the large scale structure and history of merging of galaxies and clusters. It follows that the AGWB depends on the direction of observation $\bee$,
\begin{equation}\label{e.deltaOgw}
 \Omega_{\GW}\equiv\frac{\nu_{\obs}}{\rho_c} \frac{\dd^3\rho_{\GW}(\nu_{\obs}, \bee)}{\dd\nu_{\obs}\dd^2 \bee}
 = \frac{\bar{\Omega}_{\GW}}{4\pi}+\delta\Omega_{\GW}(\bee,  \nu_{\obs})\,.
\end{equation}
Our goal is to present the properties of the anisotropic part of the AGWB. In the formalism we introduced in~\cite{Cusin:2017fwz,Cusin:2017mjm}, each galaxy is described by a GW luminosity ${\cal L}_{\Gal}(\eta,\nu_{\Gal},\theta_{\Gal})$ that depends on time ($\eta$), on the frequency in the galaxy rest frame ($\nu_{\Gal}$) and on the properties of the galaxy ($\theta_{\Gal}$, that includes e.g. its mass, metallicity). This quantity depends on all the GW sources and their galactic distribution, and therefore it represents a great source of new astrophysical information. The observed signal in a given direction is the ``sum" of the contributions over all galaxies in that direction. As it depends on the galaxy distribution, it is related to cosmology. $\delta\Omega_{\GW}(\bee,  \nu_{\obs})$ depends on the history of the large scale structure such as the mergers of galaxies and clusters but also on the initial power spectrum inherited from inflation. It is also correlated to other cosmological probes. While the theory predicts the total GW energy density in a given direction, GW detectors are sensitive to the GW strain. The link between them has been fully clarified in our former work~\cite{Cusin:2017mjm}. Our formalism, fully developed in~\cite{Cusin:2017fwz,Cusin:2017mjm}, has been applied to the case of cosmic strings in Ref. \cite{Jenkins:2018nty}.

%

We generalize  previous works~\cite{Regimbau:2011rp,Dvorkin:2016okx} in which the sources were assumed to be homogeneously and isotropically distributed, hence focusing on the monopole of the distribution. Its amplitude has been bound by LIGO~\cite{Aasi:2014zwg,TheLIGOScientific:2016dpb}, $\bar\Omega_{\GW}< 5.6\times10^{-6}$ for $\nu_{\obs}\in[41.5,169.25]$~Hz. Analysis of data from the Pulsar Timing Array (PTA) leads to $\bar\Omega_{\GW}<1.3\times10^{-9}$ for $\nu_{\obs}=2.8 \times 10^{-9}$~Hz~\cite{Shannon:2013wma}. The possibility of mapping the GW background is discussed in Refs.~\cite{Cornish:2001hg, Mitra:2007mc, Thrane:2009fp, Romano:2015uma, Romano:2016dpx, Allen:1996gp} and the reconstruction of an angular resolved map of the sky in Refs.~\cite{Cutler:1997ta, TheLIGOScientific:2016xzw}. Note that some early constraints on the anisotropy have already been obtained by PTA~\cite{Taylor2015,Sesana2008}. This is a slowly emergent field for which predictions are necessary to guide its observational prospective.

Our letter gives the first prediction of the angular power spectrum of the AGWB and its correlations with other cosmological probes. As such, this is an important prediction that opens a new field of research at the crossroad between general relativity, astrophysics and cosmology. \\

\noindent{\bf AGWB energy density.} In standard cosmology,  the universe is described by a Friedmann-Lema\^{\i}tre spacetime with perturbations that describe the large scale structure. In Newtonian gauge, its metric is $\dd s^2=a^2\left[-(1+2\Phi)\dd\eta^2+(1-2\Psi)\delta_{ij}\dd x^i\dd x^j\right]$ where the scale factor $a$ is a function of conformal time $\eta$ and $\Phi$ and $\Psi$ are the two gravitational potentials. In the theory of cosmological perturbations, any variable, $X(\eta, x^i)$ say, is a stochastic field. It can be decomposed in Fourier modes, $ X(\eta,\bk)$, which can be split as the product of a transfer function and of the initial metric perturbation as $ X(\eta,\bk)= X_k(\eta)\Phi^P(\bk)$.  The power spectrum of $\Phi^P(\bk)$ is predicted e.g. from inflation and constrained from CMB analysis (we use Planck satellite~\cite{Ade:2015xua}  cosmological parameters). Linear transfer functions are obtained from {\tt CMBquick}~\cite{CMBquick} and we use Halofit \cite{Smith:2002dz} to account for the non-linearities in the matter power spectrum. 

As any observable on the sphere, $\delta\Omega_{\GW}(\bee,  \nu_{\obs})$ can be decomposed in spherical harmonics. Its angular correlation function $\langle \delta\Omega_{\GW}(\bee_1,  \nu_{\obs})\delta\Omega_{\GW}(\bee_2,  \nu_{\obs})\rangle$ is, thanks to statistical isotropy, a function of $\cos\theta\equiv \bee_1.\bee_2$ and can be decomposed in Legendre polynomials to define the angular power spectrum $C_\ell(\nu_{\obs})$ as
\be\label{Cell}
C_{\ell}(\nu_{\obs})=\frac{2}{\pi}\int \dd k\,k^2 |{\delta\Omega}_{\ell}(k,\nu_{\obs})|^2\,.
\ee
The term ${\delta\Omega}_{\ell}(k,\nu_{\obs})$ has been derived in our previous analysis~\cite{Cusin:2017fwz}, where all the details can be found, as
\begin{widetext}
\begin{eqnarray}\label{Rkk}
{\delta\Omega}_{\ell}(k,\nu_{\obs})&&=\frac{\nu_{\obs}}{4\pi\rho_c}\int_{\eta_*}^{\eta_{\obs}} \dd\eta\, \mathcal{A}(\eta, \nu_{\obs})\left[\left(4{\Phi}_k(\eta)+b\delta_{\rm m, k}(\eta)+(b-1) 3 \mathcal{H} \frac{v_k(\eta)}{k}\right) j_{\ell}(k\Delta\eta)-2k {v}_k(\eta)j'_{\ell}(k\Delta\eta)\right]\nn\\
&&+\frac{\nu_{\obs}}{4\pi\rho_c}\int_{\eta_*}^{\eta_{\obs}} \dd\eta\,  \mathcal{B}(\eta, \nu_{\obs}) \left[-{\Phi}_k(\eta)j_{\ell}(k\Delta\eta)+k {v}_k(\eta)j'_{\ell}(k\Delta\eta)\right]\nn\\
&&+\frac{\nu_{\obs}}{4\pi\rho_c}\int_{\eta_*}^{\eta_{\obs}}
   \dd\eta\left[6\mathcal{A}(\eta,
   \nu_{\obs})-2\mathcal{B}(\eta, \nu_{\obs})\right]\int_{\eta}^{\eta_{\obs}}\dd\tilde\eta {\Phi}'_k(\tilde\eta) j_{\ell}(k\Delta\tilde\eta)\,.
\end{eqnarray}
\end{widetext}
This expression involves three types of quantities.  First, background quantities such as ${\cal H}$, the Hubble parameter in conformal time, and the look-back time, $\Delta\eta=\eta_{\obs}-\eta$. Then, the cosmological perturbations include the gravitational potential ${\Psi}_k \simeq {\Phi}_k$, the velocity field $v_k$ and the matter density $\delta_{\rm m}$. The bias $b$ relates the galaxy over-density to $\delta_{\rm m}$ and is defined as their ratio in comoving gauge, see e.g. Ref.~\cite{Bruni:2011ta}. We spot in this expression the contribution from galaxy over-density,  Sachs-Wolfe like and Doppler-like terms. Terms in the last line come from the integrated Sachs-Wolfe effect. The $j_\ell$ are spherical Bessel functions, and ${\cal A}$ and ${\cal B}$ are related to the luminosity function of a galaxy per unit of emitted frequency (${\nu}_{\Gal}\equiv (1+z) \nu_{\obs}$)
\be
\mathcal{A}(\eta, \nu_{\obs})\equiv a^4\bar{n}_{\Gal}(\eta) \int \dd\theta_{\Gal}\mathcal{L}_{\Gal}(\eta, {\nu}_{\Gal}, \theta_{\Gal})\,,\label{AA}
\ee
\be
\mathcal{B}(\eta, \nu_{\obs})\equiv a^3\,\nu_{\obs}\bar{n}_{\Gal}(\eta)\int \dd\theta_{\Gal}\frac{\partial \mathcal{L}_{\Gal}}{\partial {\nu}_{\Gal}}(\eta,  {\nu}_{\Gal}, \theta_{\Gal})\,.\label{BB}
\ee

\noindent{\bf Astrophysical model.}  The computation of $\mathcal{A}$ and $\mathcal{B}$ requires (1) the luminosity $\mathcal{L}_{\Gal}(z, \nu_{\Gal},  M_{\Gal})$ of a galaxy with halo mass $M_{\Gal}$ at redshift $z$ and (2) to sum it over the entire galaxy population using the halo mass function $\dd n/\dd M(M,z)$ calibrated to numerical simulations~\cite{2008ApJ...688..709T}. Our computation follows the formalism we developed in Refs.~\cite{Dvorkin:2016okx,Dvorkin:2016wac,2017arXiv170909197D}. It has three main steps: (1) the determination of  the emitted energy spectrum $\dd E/\dd\nu$ of each type of GW sources, (2) the evolution of the population of this type of sources in a galaxy of mass $M_{\Gal}$ and (3) the integration over the galaxy population.

Galaxies contain many sources of GW. We concentrate on the background from 
merging binary BHs which contribute to the range of frequencies between a few Hz to a few hundreds of Hz. We describe the emitted energy using the results of Ref.~\cite{2008PhRvD..77j4017A}.

Let us turn to the computation of the number of BH mergers per unit time in a galaxy of mass $M_{\Gal}$. 

First, we calculate the formation rate of binary BHs, $\mathcal{R}_f$. It depends on the star formation rate (SFR) $\psi(M_{\Gal},t)$, given in units of $M_{\odot}/{\rm yr}$  and on the evolution and lifetime of these stars, which is determined by their initial mass and metallicity. We use the SFR of  Ref.~\cite{2013ApJ...770...57B} to calculate the mass in stars of a galaxy. It provides a fit to a large number of observables, including the stellar mass function, the specific SFR and cosmic star formation rates from $z=0$ to $z=8$ and in the halo mass range of $10^9-10^{15}M_{\odot}$. Then, we use the Salpeter initial mass function (IMF)~\cite{1955ApJ...121..161S} to describe the number of stars per unit total stellar mass formed, $\phi=\dd N/\dd M_*\dd M_{\rm tot,*}\propto M_*^{-2.35}$, where $M_*$ is the mass of the star at birth.

Second, we need to describe the evolution of a star of initial mass $M_*$. Its lifetime $\tau(M_*,Z_*)$ and the end point (BH or NS) of its evolution depend on both $M_*$ and its  metallicity. We use the stellar evolution model by Ref.~\cite{2012ApJ...749...91F} to obtain the function $m=g_s(M_*,Z)$ that gives the mass of the BH formed for a star with initial $(M_*,Z)$ and Ref.~\cite{2002A&A...382...28S} to calculate stellar lifetimes $\tau(M_*,Z_*)$. Typically, massive stars ($M_*\gtrsim8M_\odot$) explode as supernovae or collapse to form BH on a timescale of a few Myr. If we assume such short stellar lifetimes, the stellar metallicity tracks the metallicity of the interstellar medium (ISM) given by $Z=Z(M_{\Gal},z)$. We adopt the prescription of Ref.~\cite{2016MNRAS.456.2140M} for the ISM metallicity as a function of galaxy mass and redshift.

\begin{figure}
\centering
\includegraphics[width=\columnwidth]{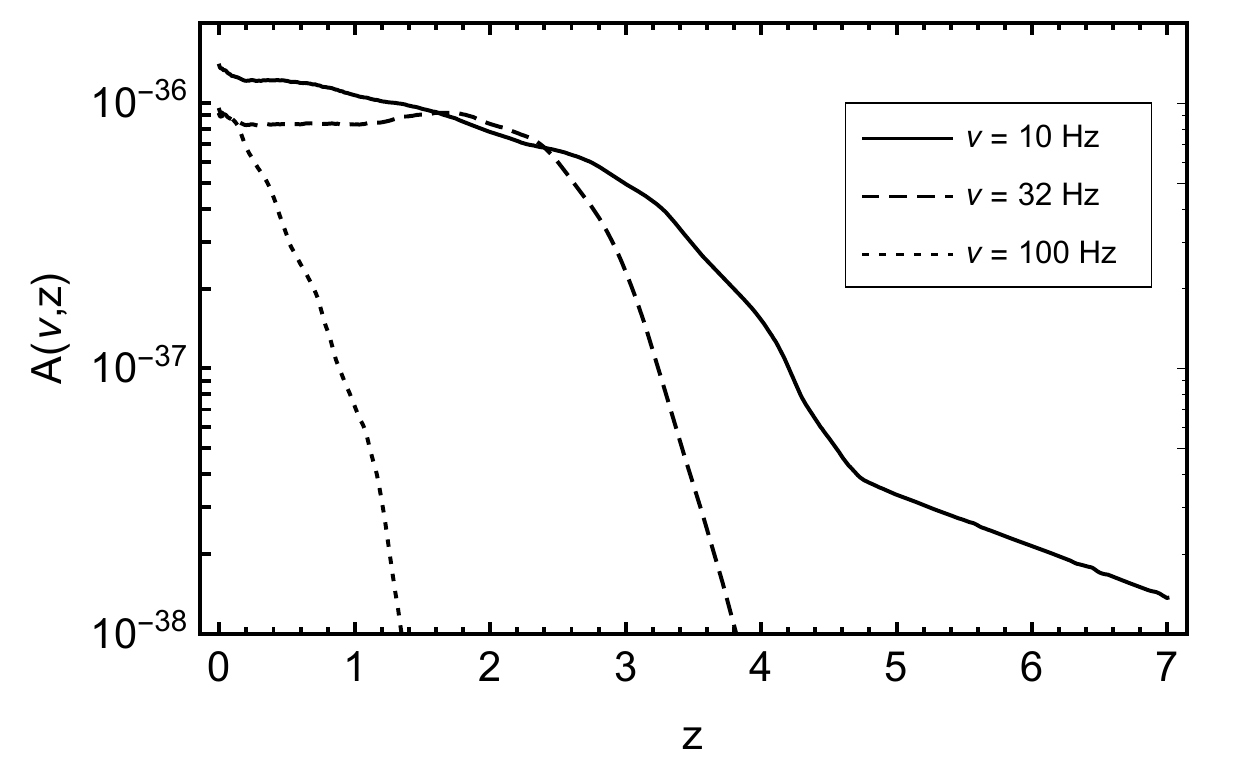}
\caption{The astrophysical source function $\mathcal{A}(\nu_{\obs}, z)$ as a function of redshift, for different frequency bands.}\label{FIG.anuz}
  \end{figure}

Under these assumptions, the instantaneous rate of BH formation at a given cosmic time $t$  and for a galaxy $M_{\Gal}$ in units of events per unit BH mass $m$ is given by ${\cal R}_1(m,t)=\psi[M_{\Gal},t-\tau(M_*)] \phi(M_*)\times \dd M_*/\dd m$ where $M_*(m)$ and $\dd M_*/\dd m$ are deduced from the relation $m=g_s(M_*,Z)$. We then assume that only a fraction $\beta=0.005$ of these BHs resides in binary systems  so that the rate of formation of the latter is ${\cal R}_2(m,t)=\beta {\cal R}_1(m,t)$ (chosen to match the mean GW density in Refs.~\cite{Dvorkin:2016okx,Dvorkin:2016wac}). And, following Ref.~\cite{2017arXiv170909197D}, the birth rate of binaries with component masses $(m,m' \leq m)$ is ${\cal R}_{\rm bin}(m',m)= {\cal R}_2(m){\cal R}_2(m')P(m,m')$ where $P(m', m)$ is the 2-mass distribution of binaries. We assume $P(m, m')=$~cnst in the following.

The merger rate is obtained once the time to coalescence of the binaries is known. This requires the distribution of the orbital parameters $(a_{\rm f},e_{\rm f})$ at the time of formation. Since orbits circularize quickly, we assume zero eccentricity for all the binaries. We also assume that the distribution of the semi-major axis \emph{at formation} is $f(a_{\rm f})\propto a_{\rm f}^{-1}$ with cut-off at $a_{\rm f,min}=0.2$ AU and $a_{\rm f,max}=5000$ AU \citep{Dvorkin:2016okx}. Hence, the birth rate of BH binaries (per unit mass squared per unit time and per units of $a_{\rm f}$) is $\mathcal{R}_f[m,m', a_{\rm f}, t]={\cal R}_{\rm bin}(m',m)f(a_{\rm f})$ from which we deduce that the merger rate at time $t$ is  $\mathcal{R}_{\m}[m, m', a_{\rm f}, t]= \mathcal{R}_f[m, m', a_{\rm f}, t-\tau_{\rm m}(m, m', a_f)]$ where $\tau_{\rm m}(m, m', a_{\rm f})$ is the merger time of the system $(m,m', a_{\rm f})$.

The GW luminosity is the overall contribution of the mergers, i.e.
\be
 \mathcal{L}_{\Gal}=\int \dd m\, \dd m'\, \dd a_{\rm f}\, \frac{\dd E}{\dd \nu}\times \mathcal{R}_{\m}[m, m', a_{\rm f}, t]\,.
\ee
Its integration~(\ref{AA})  on $\theta_{\Gal}$ reduces to an integration on $M_{\Gal}$, hence weighted by the halo mass function and gives ${\cal A}$ presented on Fig.~\ref{FIG.anuz}. We check that $\nu_{\obs}\int{\cal A}(\nu_{\obs},\eta)\dd\eta/\rho_c$ matches with the mean GW density $\bar\Omega_{\GW}(\nu_{\obs})$ computed in Refs.~\cite{Dvorkin:2016okx,Dvorkin:2016wac}, where the BH merger rate was normalised to the observed rates.\\

\begin{figure}
\centering
\includegraphics[width=\columnwidth]{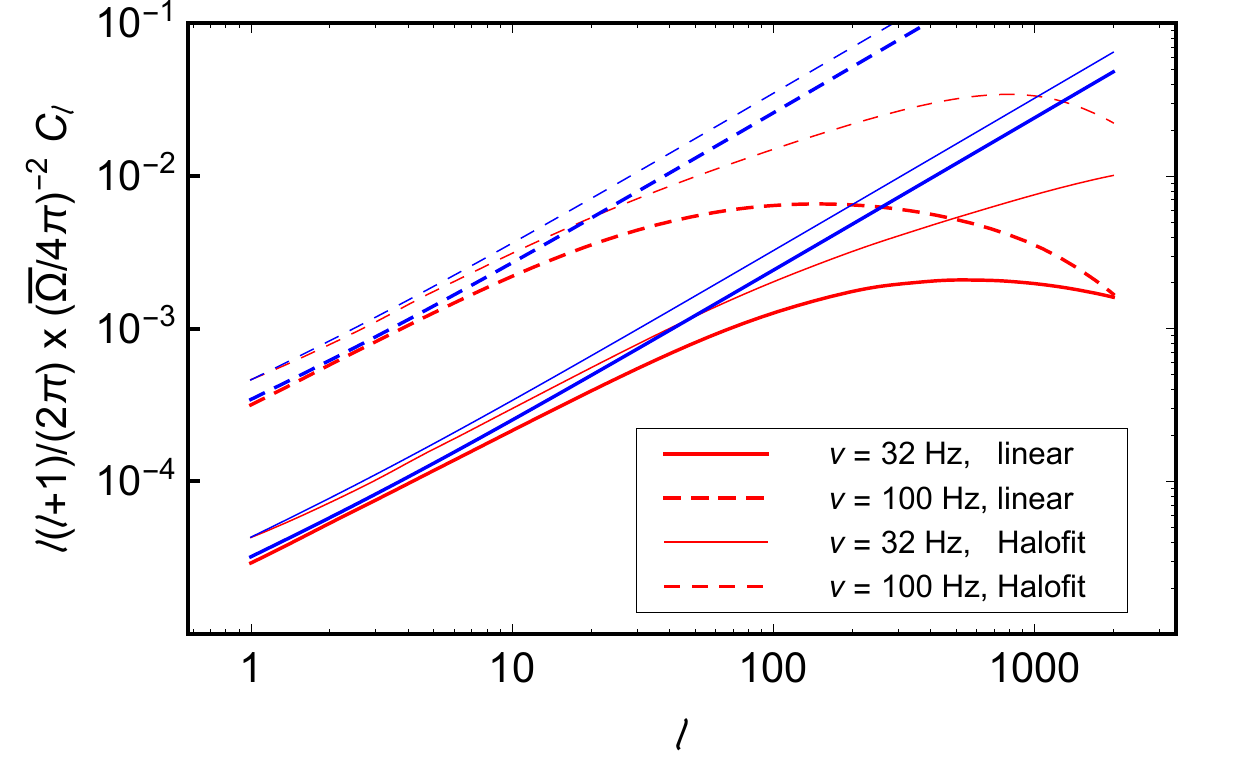}
 \caption{Angular power spectrum of AGWB density fluctuation normalized to the monopole, for linear and non-linear (Halofit) spectra of density fluctuations. The straight blue lines are the associated large scale approximations~(\ref{ClLargeScale}).}\label{FIG.Cl}
  \end{figure}

\noindent{\bf Angular power spectrum.} Figure~\ref{FIG.Cl} presents the $C_\ell$'s for different frequency bands. We estimate in Ref.~\cite{PRD}  that on large scales, we have approximately 
\be\label{ClLargeScale}
(\ell+\tfrac{1}{2}) C_\ell(\nu_{\obs}) \simeq \left[\frac{\nu_{\obs}\mathcal{A}(\eta_{\obs},\nu_{\obs}) b(\eta_{\obs}) }{4\pi \rho_{\rm c} } \right]^2 \int_{k_{\rm min}} P_\delta(k) \dd k\,,
\ee 
where $P_\delta(k) $ is the matter power spectrum today, $k_{\rm min} \equiv 1/\eta_{\obs}$ and we are considering a scale-independent bias $b\propto \sqrt{1+z}$; see e.g. \cite{Marin:2013bbb, Rassat:2008ja}. Clearly, non-linearities shift the amplitude but do not alter significantly the shape of the angular power spectrum. They  affect both small and large angular scales because the function ${\cal A}$ extends up to today, in contrast with what happens e.g. for the CMB, or galaxy surveys when considering a given redshift bin.  The variance of $\Omega_{\rm GW}$ due to the distribution of the large scale structures is
\be
 \sigma^2_{\GW}(\nu_{\obs})\equiv \sum_{\ell}\frac{(2\ell+1)}{4\pi}C_{\ell}(\nu_{\obs})\,.
\ee
Considering multipoles up to $\ell_{\rm max} = 2000$, we find $4\pi\sigma_{\rm GW} /\bar \Omega_{\GW} \simeq 0.14$ (resp. $0.32$) for $\nu_{\obs}=32\,{\rm Hz}$ (resp. $\nu_{\obs} = 100\,{\rm Hz}$), that is variations of the AGWB are typically of order $14\%$ (resp. $32\%$) at $32\,{\rm Hz}$ (resp. $100\,{\rm Hz}$). When using only the linear power spectrum, these values are approximately halved.\\

\begin{figure}\label{FIG.cross}
\centering
\includegraphics[width=\columnwidth]{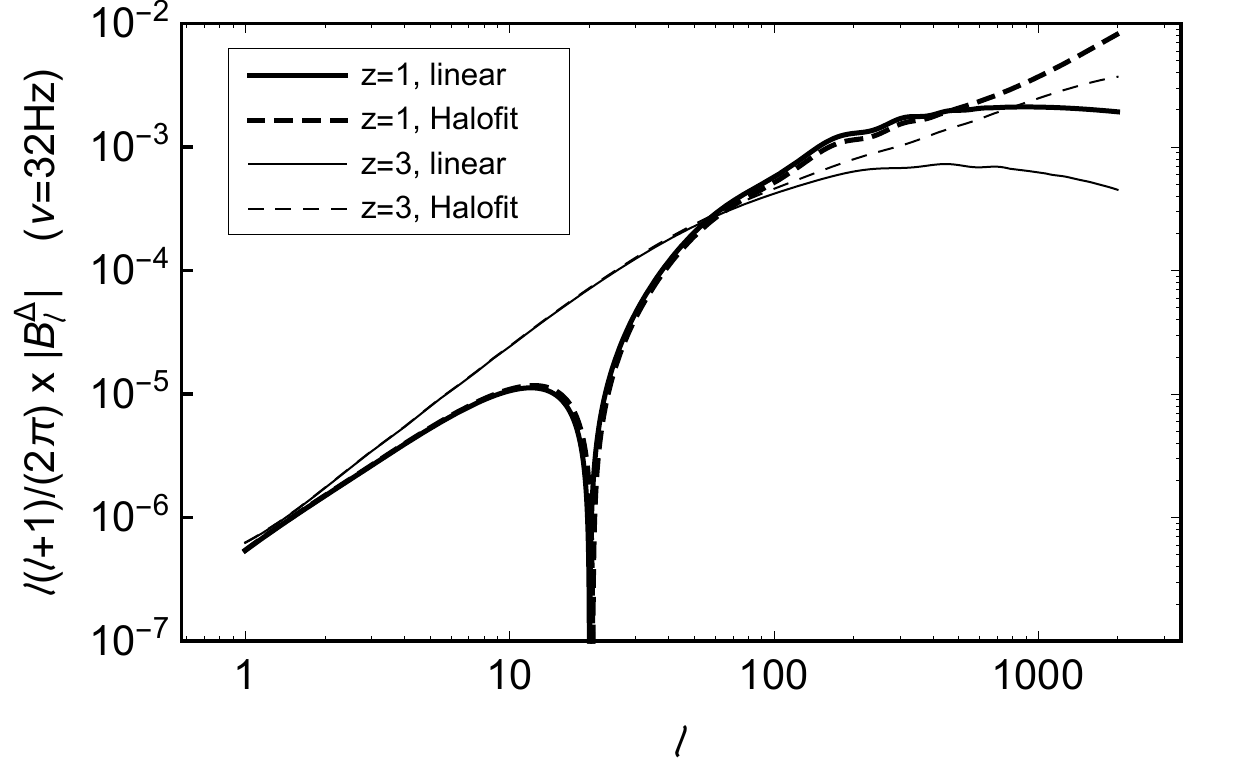}\\
 \includegraphics[width=\columnwidth]{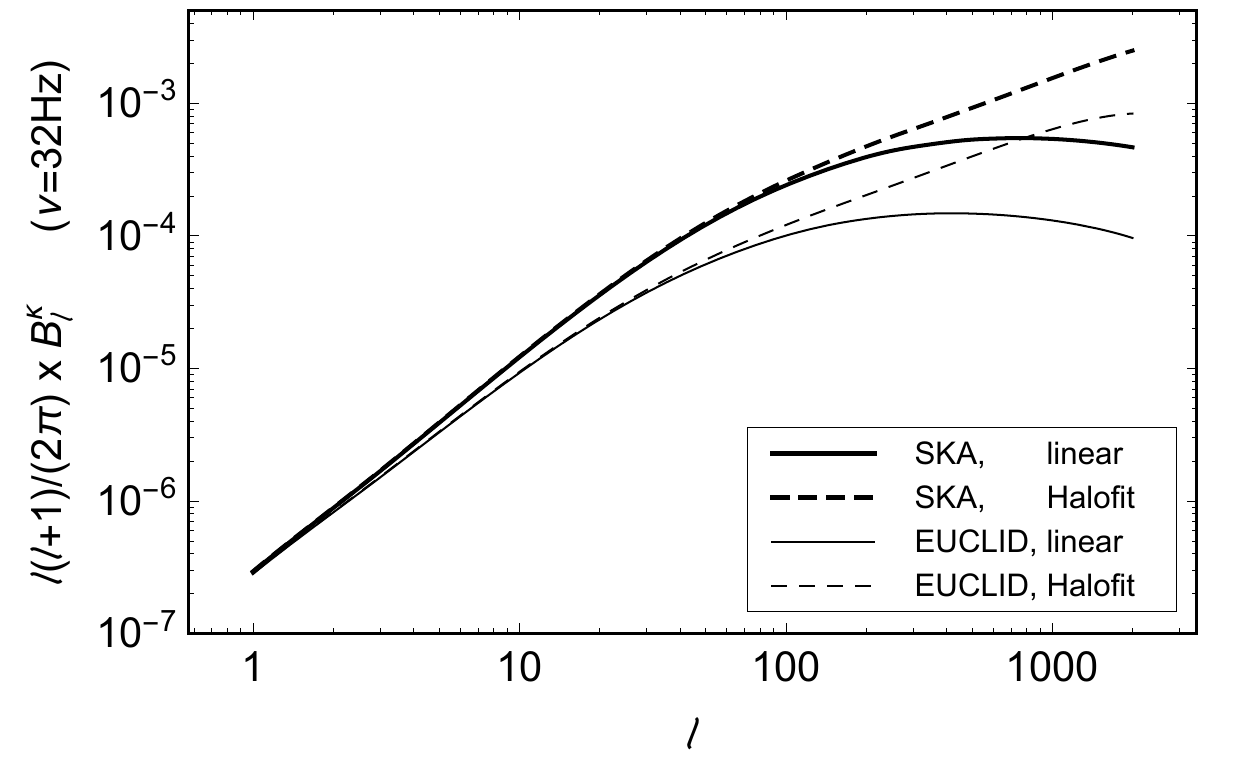}
    \caption{\label{crosscorr} {\it Top} : cross-correlation with galaxy number counts. {\it Bottom} :  cross-correlations with convergence ($\kappa$) using the SKA~\cite{Andrianomena:2014sya} and Euclid~\cite{2011arXiv1110.3193L} redshift distributions.}
  \end{figure}

\noindent{\bf Cross-correlations.} Since Eq.~(\ref{Rkk}) depends on cosmological perturbations, it correlates with any other cosmological probe, as galaxy number counts and weak lensing convergence. The cross-correlation power spectra have been computed in Ref.~\cite{Cusin:2017fwz}, 
\be
B_\ell^X(\nu_{\obs}) \equiv \frac{2}{\pi}\int \dd k\,k^2 \frac{4\pi}{\bar \Omega_{\GW}(\nu_{\obs})} \delta\Omega^*_{\ell}(k,\nu_{\obs})\, X_{\ell}(k)\,,
\ee
with $X_\ell=\kappa_\ell$ for cosmic convergence given by Eq.~(100) of Ref.~\cite{Cusin:2017fwz} and $X_\ell=\Delta_\ell$ for galaxy number counts  given by Eq.~(129) of Ref.~\cite{Cusin:2017fwz} in the Kaiser approximation, on which we also add the effect of weak-lensing convergence as in Eq. (13) of Ref.~\cite{Bruni:2011ta}. They are depicted in Fig.~\ref{crosscorr}.\\

\noindent{\bf Conclusions and perspectives.} This letter has presented the first numerical computation of the AGWB power spectrum and its correlations with galaxy number counts and weak lensing following the formalism we developed in Refs.~\cite{Cusin:2017fwz,Cusin:2017mjm}. These spectra depend on cosmology through the transfer functions and the initial power spectrum and on astrophysics through the merger history of galaxies, the SFR, IMF and stellar evolution that determine the mass distribution of BH and NS. The GW luminosity function also depends on the distribution of the initial orbital parameters of the binaries, which influences their lifetime and on general relativistic models for the emitted energy of each type of sources.  Indeed, many ingredients of this calculation (such as the distribution of the orbital parameter, the function $P(m,m')$, the contribution of other GW sources) come with an associated uncertainty  and for this study we have adopted standard prescriptions. The effect of these choices on the monopole of the AGWB was discussed in Ref.~\cite{Dvorkin:2016okx} and a full analysis of their impact, as well as the one of the cosmological parameters and of the non-linear regime, on the AGWB power spectra will be presented in a companion article~\cite{PRD}.

This analysis has already determined the general shape of the power spectrum and the variance of $\delta \Omega_{\GW}$. To get closer to observations, one will also need to design estimators of this spectra and evaluate their signal-to-noise ratio for forthcoming experiments. 
The observation of the AGWB and its anisotropies will convey us information about both astrophysics and cosmology. In particular, comparing predictions to observations will allow to put new types of constraints on astrophysical models. 
The study of the AGWB can impact astrophysics as much as CMB did for cosmology during the past decades.\\


\noindent{\bf Acknowledgements} 
GC thanks IAP for hospitality during the first stage of this work. We thank Camille Bonvin, Elisabeth Vangioni, Tania Regimbau, Joseph Romano, Alberto Sesana and Bernard Whiting for stimulating discussions. This work was done within the Labex ILP (reference ANR-10-LABX-63), part of the Idex SUPER, and received financial state aid managed by the ANR, as part of the programme Investissements d'avenir under the reference ANR-11-IDEX-0004-02. We acknowledge the financial support from the EMERGENCE 2016 project, Sorbonne Universit\'es, convention no. SU-16- R-EMR-61 (MODOG). GC acknowledges financial support from ERC Grant  No: 693024 and 
Beecroft Trust. 


\bibliography{BH+BH-refs}

\end{document}